\documentclass[a4paper,fleqn,11pt]{article}
\usepackage{graphicx,array}
\usepackage{amssymb,amsmath}
\usepackage{authblk}
\renewcommand{\normalsize}{\fontsize{9pt}{15pt}\selectfont}
\newcommand{\halfnormalsize}{\fontsize{9pt}{12pt}\selectfont}
\renewcommand\title[1]{
\begin{center}\fontsize{12pt}{15pt}{\bf{\bfseries\sffamily#1}\par}\end{center}}
\renewcommand\author[1]{\begin{center}\small{\sffamily#1}\end{center}\par}
\renewcommand\abstract[1]{\begin{quote}\footnotesize #1\end{quote}\par}
\makeatletter
\renewcommand{\fnum@figure}[1]{\fontsize{8pt}{12pt}{\textbf{Figure~\thefigure. }}}
\renewcommand{\fnum@table}[1]{\fontsize{8pt}{12pt}{\textbf{Table~\thetable. }}}
\@ifundefined{captionstyle}{}{}
\@ifundefined{instindent}{\newdimen\instindent}{}
\long\def\@caption#1[#2]#3{\par\addcontentsline{\csname
  ext@#1\endcsname}{#1}{\protect\numberline{\csname
  the#1\endcsname}{\ignorespaces #2}}\begingroup
    \@parboxrestore\if@minipage\@setminipage\fi
    \@makecaption{\csname fnum@#1\endcsname}{\ignorespaces #3}\par
  \endgroup}
\renewcommand\@seccntformat[1]{\csname prefmt@#1\endcsname
	\csname the#1\endcsname. \csname postfmt@#1\endcsname}
\newcommand\postfmt@section{\hskip 1mm}
\newcommand\postfmt@subsection{\hskip 1mm}
\renewcommand\section{\@startsection{section}{1}{\z@}%
{-12pt plus3pt minus2pt}%
{3pt}%
	{\normalfont\normalsize\bfseries}}
\renewcommand\subsection{\@startsection{subsection}{2}{\z@}%
{-9pt plus3pt minus2pt}%
{3pt}%
	{\normalfont\normalsize\bfseries}}
\renewcommand\subsubsection{\@startsection{subsubsection}{3}{\z@}%
{-9pt plus3pt minus2pt}%
{3pt}%
	{\normalfont\normalsize\bfseries}}
\makeatother
\newcommand\stamp[6]{\vspace{0.25cm}\linespread{1}\noindent\footnotesize
#1,\ #2:\ #3,\ #4 ({\em #5}).\ #6\par}
\def\({\left(}
\def\){\right)}

\begin{document}
\vspace*{5.4 mm}

\title{Quantum-gravity in a dynamical system perspective}
\vspace*{3.8 mm}

\author{Sijo K. Joseph}
%\affil{WellGreen Technologies Pvt. Ltd., Thodupuzha, Kerala, India}

\abstract{{\em Abstract:} General Theory of Relativity and Quantum theory gives two different
description of the same mother nature in the big and small scale respectively.
Mathematical languages of these two theories are entirely different, one is geometric
while the other one is probabilistic. This curious feature makes the merging of these
theories (quantum-gravity) considerably difficult. In this manuscript, we explore
quantum-gravity in a dynamical system perspective. For example, in the standard
quantum theory, the wave equation is a linear partial differential equation while in
General Theory of relativity, the field equations are highly nonlinear. A classical
dynamical system can show very rich phase-space structures which is absent in a
linear partial differential equation. In order to incorporate gravitational corrections
one can think about the nonlinear extensions of quantum theory. In this manuscript
such an attempt to quantum gravity is explored. Dissipative and forcing corrections
are found in the newly formulated quantum equation and it's physical interpretations
are given.}

\section{Introduction}
It is reasonable to argue that the linear quantum theory might be a theoretical approximation which arises 
from a more fundamental nonlinear theory. Following this idea, many researchers had performed their 
analysis and different extensions of the Schr\"odinger equation are obtained~\cite{Doebner1992,Svetlichny2005,Castro:2006}. It is well known that 
in the many-body quantum theory, one needs to deal with nonlinear equations like the Kohn-Sham equation~
\cite{KohnSham1965}. Similarly in the case of Bose-Einstein condensate one needs to deal with a nonlinear Gross-Pitaekvskii equation~\cite{Gross1961,Pitaevskii1961}. There had been attempts to generalize the fundamental quantum mechanics laws to nonlinear regime. 
There are several problems related to this, for example the energy functional computed using the nonlinear 
Hamiltonian seems to give a different value compared to the quantum field theory calculation. But still 
there are a special classes of nonlinear modification which will give the same results~\cite{Puszkarz1998}. T. P. Singh et al. had demonstrated that the nonlinear quantum mechanics is necessary, once the quantum mechanical space-time background is taken into account~\cite{TPSingh2005}.  

Apart from these line of arguments, in quantum chaos a purely philosophical issue appears. It is 
considered that quantum theory is more fundamental and classical mechanics appears as a particular 
limiting case. In orthodox quantum theory, field equations are linear but in the classical limit, the 
dynamics of the corresponding classical system can be nonlinear and it can show very rich phase-space 
structures. In the sense of this rich phase-space structure, How come quantum theory is mathematically a 
subset of classical mechanics ? This is a serious conceptual issue, but in quantum chaos community this is 
not taken seriously as a conceptual one. Quantum system is inherently linear, hence chaotic property is 
absent there. One is compelled to study the manifestation of chaos in a quantum system. Usually the energy 
level statistic is studied to analyze the quantum signature of chaos~\cite{Bohigas1984,Gutzwiller1990}, while some of the research  tends to explore the quantum entanglement to find the quantum signature of chaos~\cite{sijo_physlettA,sijo_optiexp,SKJ_Entropy2016,SKJ_IJBC2016}.
Quantum theory should contain all the mathematical property to describe the reality completely. In other 
words, if quantum theory is more fundamental, it should be able to exhibit all the mathematical 
properties of a classical system. For example, a classical system can have a chaotic behavior while a 
quantum system can only show the signatures of the classical chaos; this should be seen as a theoretical 
limitation of the orthodox quantum theory. Another conceptual issue arises when one tries to combine quantum 
theory with general theory of relativity. One theory consists of real and nonlinear field equation while the other one is linear with complex field equation and combining them in a consistent manner is still a tedious task.

These issues can be circumvented by adopting the 
deBroglie-Bohm (dBB) version of the quantum  theory. In dBB version of quantum physics, field equations are nonlinear and it is free from complex quantities. More importantly, in this formalism  particles and waves exists simultaneously\cite{BohmI,BohmII} and the particle can exhibit chaotic dynamics. 
Extending dBB version of quantum theory to the relativistic regime as explored by Shojai 
et al.~\cite{Shojai_Article}, one can immediately see that the relativistic quantum theory in dBB formalism is equivalent to a 
conformally flat general theory of relativity. Quantum trajectories in the dBB version can be seen as a 
geodesic of the particle on a curved manifold. Trajectories in de-Broglie Bohm theory is no longer 
surreal, a definite physical meaning is attached to it.  Relativistic dBB 
interpretation can be stated in a different manner, it is equivalent to the Einstein's General theory of 
relativity. The pilot-wave determines the space-time manifold structure and the particles trajectories 
(Bohmian trajectories) are the geodesic of this curved manifold.

\section{Probability vs Geometry}
%Orthodox quantum mechanics has two interesting features, a state vector is time evolved by a unitary 
%operator and the measurement can collapse a state vector. There is always a concern about whether the wave-
%function is physical entity. Wavefunction $\psi$ only describe an amplitude, experimentally observable 
%quantities are related to the square of the wave-function $|\psi|^2$. In quantum physics observable 
%quantities are always intermingled with quantum probabilities. The space-time in usual quantum mechanics 
%is Minkowskian $(\eta_{\mu\nu})$ and the mathematics seems to be entirely different than Einstein's 
%general theory of relativity.  

In Einstein's general theory of relativity, geometry of the space-time plays an important role. 
Particles follows the geodesic on a curved manifold and the energy-momentum tensor of an object 
curves the space-time. It is curious to observe that when the same universe is modeled in big scale 
and small scale we obtained two theories, while the mathematical structure of these two theories are highly different. 
In order to merge these two theories (general relativity and quantum theory), one needs to think about how these theories are different. 
Is there any mathematical parallelism between these two theories? 
One is modeling the same universe in the big scale and small scale, and one can get two different mathematical description. If these mathematical descriptions are individually complete, there must be a duality between these two theories. Our choice of particular mathematical tools may give us one theory or the other. In this perspective, there must exist a possibility that relativistic quantum physics should be expressible in terms of general relativity and vice versa. 

Following this line of thought and considering the scalar matter field, ie. the Klein-Gordon equation on a flat space-time, which is given by,
\begin{eqnarray}
\Bigl(\Box+\frac{m^2}{\hbar^2}\Bigr)\phi=0.
\end{eqnarray}
As pointed out by F. Shojai~\cite{Shojai_Article}, one can think about the Bohmian representation of this equation by taking $\phi=\sqrt{\rho}e^{\frac{i}{h}S}$ and can obtain the following coupled equations,
\begin{eqnarray}
\partial_{\mu}(\rho\partial^{\mu}S)&=&0 \label{ContiEqflat} \\
\partial_{\mu}S \partial^{\mu}S&=& m^{2} \Omega^2,  \label{HJEqflat}
\end{eqnarray}
where $\Omega^2=1+\frac{\hbar^2}{m^2}\frac{\Box\sqrt{\rho}}{\sqrt{\rho}}$.
This particular form of Eq.~\ref{ContiEqflat} and Eq.~\ref{HJEqflat} is highly interesting due to its easiness to 
incorporate it with Einstein's General Theory of Relativity. On a curved manifold, these coupled equations can be easily generalized by changing the usual derivative $\partial_{\mu}$ to covariant one $\nabla_{\mu}$. Still these modifications can only incorporates gravitational effects on the quantum mechanical matter field. The continuity equation doesn't contain any contribution from the conformal factor $\Omega^2$.
Hence one needs to correct these equations to fully describe the interaction between gravitational field
and the 
scalar quantum-mechanical matter field. Once we incorporate the matter Lagrangian with conformally
transformed Einstein-Hilbert action, these corrections will naturally appear. 
%\begin{eqnarray}
%\nabla_{\mu}(\rho\Omega^2 \nabla^{\mu}S)=0 \\
%\nabla_{\mu}S \nabla^{\mu}S= m^{2} \Omega^2,  \label{HJeq}
%\end{eqnarray}
%where $\Omega^2= \exp{(\frac{\hbar^{2}}{m^2} \frac{\Box^{2}\sqrt{\rho}}{\sqrt{\rho}})}$.
Returning to Eq.~\ref{HJEqflat}, it is just the Hamilton-Jacobi equation with some quantum correction 
and it can be rewritten as,

\begin{eqnarray}
\frac{1}{\Omega^2}\eta^{\mu \nu}\partial_{\mu}S \partial_{\nu}S=m^{2}. \label{geometriceq}
\end{eqnarray}

As pointed out by Shojai et al.~\cite{Shojai_Article}, this is the free particle equation of motion on a curved manifold with a metric $g_{\mu \nu}$. 
Since $g^{\mu \nu}=\frac{1}{\Omega^2}\eta^{\mu \nu}$, the metric tensor of the quantum mechanical manifold is given by.
\begin{eqnarray}
{g}_{\mu \nu}=\Omega^2  \eta_{\mu \nu} \label{metricqm}
\end{eqnarray}
This implies ${g}_{\mu \nu}=\Omega^2  \eta_{\mu \nu}$ where $\Omega^2=1+\frac{\hbar^2}{m^2}\frac{\bar{\Box}\sqrt{\rho}}{\sqrt{\rho}}=1+Q$. For a more general theory one can take $\Omega^2=\exp{(Q)}$, but in this case we are taking the contribution of the higher powers of $Q$ in the conformal factor. One should note that, in this simple generalization of conformal factor, the quantum theory is fundamentally modified~\cite{GabayJoseph2}.

Here Eq.~\ref{geometriceq} and Eq.~\ref{metricqm} simply conveys that the quantum effects can be simply absorbed into the curved space-time. The flat space-time $\eta_{\mu \nu}$ appears since the effect of gravity is not considered. If one needs to incorporate gravity, it is pretty straight forward to assume a curved background metric instead of $\eta_{\mu \nu}$. Assuming that the effect of gravity exist, then the metric due to the gravitational field is denoted by $\bar{g}_{\mu \nu}$. Now the effective metric of the space-time including the gravitational field contribution is given by,
\begin{eqnarray}
g_{\mu \nu}=\Omega^2 \bar{g}_{\mu \nu}
\end{eqnarray}
%where $\Omega^2=1+\frac{\hbar^2}{m^2}\frac{\bar{\Box}\sqrt{\rho}}{\sqrt{\rho}}$.
%Now the equations of motion is given by,
%\begin{eqnarray}
%\nabla_{\mu}S \nabla^{\mu}S=m^2 \label{FreeEqMot}
%\end{eqnarray}
It is already shown that the Eq.~\ref{geometriceq} simply indicates that, 
the quantum particle moves like a free particle on a curved manifold described by the metric $g_{\mu \nu}$.This is true even in the presence of the gravitationally curved background space-time $\bar{g}_{\mu \nu}$.
From the very basic definition of the Christoffel symbol,
\begin{eqnarray}
\Gamma^i{}_{k\ell}=\frac{1}{2}g^{im} \left(\frac{\partial g_{mk}}{\partial x^\ell} + \frac{\partial g_{m\ell}}{\partial x^k} - \frac{\partial g_{k\ell}}{\partial x^m} \right) 
\end{eqnarray}

Substituting $g_{\mu \nu}=\Omega^2 \bar{g}_{\mu \nu}$ into the above expression,
\begin{eqnarray}
\frac{\partial g_{mk}}{\partial x^\ell} = \Omega^2  \frac{\partial \bar{g}_{mk}}{\partial x^\ell} +  \bar{g}_{mk} \frac{\partial \Omega^2}{\partial x^\ell}.
\end{eqnarray}
Hence the Christoffel symbols becomes,
\begin{eqnarray}
\Gamma^i{}_{k\ell}={\bar{\Gamma}^i{}_{k\ell}}
+ \frac{1}{2\Omega^2} \left(\delta^i{}_{k} \frac{\partial \Omega^2}{\partial x^\ell} +
\delta^i{}_{\ell} \frac{\partial \Omega^2}{\partial x^k} - 
\bar{g}^{im} \bar{g}_{k\ell} \frac{\partial \Omega^2}{\partial x^m} \right) \nonumber \\ \label{qchrist}
\end{eqnarray}
In general theory of relativity, the geodesic equation is given by,
\begin{eqnarray}
\frac{d^2 x^i }{d\tau^2} + {\Gamma}^i{}_{k\ell} \frac{dx^k }{d\tau}\frac{dx^\ell }{d\tau} = 0
\end{eqnarray}
Considering the quantum effect, the Christoffel symbol is computed using Eq.~\ref{qchrist} and substituting it in the geodesic equation, we get,
\begin{eqnarray}
\frac{d^2 x^i }{d\tau^2} +\Bigg({\bar{\Gamma}^i{}_{k\ell}}+{{\Gamma_{\bf{qm}}}}^i{}_{k\ell}\Bigg) \frac{dx^k }{d\tau}\frac{dx^\ell }{d\tau} = 0,
\end{eqnarray}
where ${\Gamma_{\bf{qm}}}^i{}_{k\ell}= \frac{1}{2\Omega^2} \left(\delta^i{}_{k} \frac{\partial \Omega^2}{\partial x^\ell} + 
\delta^i{}_{\ell} \frac{\partial \Omega^2}{\partial x^k} -
\bar{g}^{im} \bar{g}_{k\ell} \frac{\partial \Omega^2}{\partial x^m} \right)$ is the quantum mechanical Christoffel symbol.
This general expression for the geodesic equation is not a surprise due to the fact that a quantum force is present in the dBB interpretation of quantum mechanics. Quantum Christoffel symbol arises, since we are in a frame where the metric is curved due to the quantum mechanical force. 

In dBB formulation of quantum mechanics, the quantum force arises from the quantum potential 
$Q=\frac{\hbar^{2}}{m^2} \frac{\Box^{2}\sqrt{\rho}}{\sqrt{\rho}}$.
This quantum potential is intimately connected to the conformal factor $\Omega^{2}=e^Q$. Note that the geodesic equation can be entirely
written in-terms of quantum mechanical density $\rho$. Here the Quantum Christoffel symbol arises from an underlying quantum mechanical wavefunction. In the usual principle of equivalence, the gravitational potential is formulated in a geometrical way. 
In the quantum regime one cannot ignore the effect of the quantum potential $Q$ and its curvature contribution,
this will result into the generalized equivalence principle. 
It can be easily seen that, as $\hbar\to 0$ the quantum effect becomes negligible which results $\Omega^2\approx1$. 
Hence the contributions to the Christoffel symbols due to the space-time variation of $\Omega^2$ becomes negligible, hence we get

\begin{eqnarray}
\frac{d^2 x^i }{d\tau^2} +{\bar{\Gamma}^i{}_{k\ell}} \frac{dx^k }{d\tau}\frac{dx^\ell }{d\tau} = 0.
\end{eqnarray}

It can be easily seen that we recover the standard geodesic equation in the general relativity.
In the classical limit of the equation, only gravitational field is taken into account and ${\bar{\Gamma}^i{}_{k\ell}}$ 
denote the Christoffel symbol due to the gravitational field only. On the other hand, 
a pure quantum version of the Christoffel symbol can be found by taking $\bar{g}_{\mu \nu}=\eta_{\mu \nu}$. 
Then the geodesic equation for a free quantum particle can be written as,

\begin{eqnarray}
\frac{d^2 x^i }{d\tau^2}+\frac{1}{2\Omega^2}\left(\delta^i{}_{k}\frac{\partial\Omega^2}{\partial x^\ell} 
+\delta^i{}_{\ell}\frac{\partial \Omega^2}{\partial x^k}-{\eta}^{im}{\eta}_{k\ell} \frac{\partial\Omega^2}{\partial x^m}\right) \frac{dx^k }{d\tau}\frac{dx^\ell }{d\tau}=0. \nonumber \\
\end{eqnarray} 

In terms of the quantum potential, the quantum mechanical geodesic equation is written as follows,

\begin{eqnarray}
\frac{d^2 x^i }{d\tau^2}+\frac{1}{2}\left(\delta^i{}_{k}\frac{\partial Q}{\partial x^\ell} 
+\delta^i{}_{\ell}\frac{\partial Q}{\partial x^k}-{\eta}^{im}{\eta}_{k\ell} \frac{\partial Q}{\partial x^m}\right) \frac{dx^k }{d\tau}\frac{dx^\ell }{d\tau}=0, \nonumber \\
\end{eqnarray} 
where $Q=\frac{\hbar^{2}}{m^2} \frac{\Box\sqrt{\rho}}{\sqrt{\rho}}$ and $\Omega^2=\exp{(Q)}$ is assumed. Here quantum mechanical Christoffel symbol purely depends on the 4-dimensional density distribution of the quantum particle and the quantum mechanical acceleration depends on the derivative of the quantum potential $Q$  with respect to the space-time variables (quantum force).

A theory which is based on purely complex quantities are hard to interpret while a geometrical theory is comparably easy to interpret. It is to be noted that the Bohmian interpretation of quantum mechanics has some nice features of classical mechanics. But the quantum potential makes the things complicated.
The presence of the guiding equation is a peculiar features which makes dBB theory different from a pure classical theory. The guiding equation relates particle velocity to the quantum mechanical wave-function.
This can be understood in a more clear way in a relativistic dBB theory. Particle momentum is determined by the local structure of the space-time manifold. The guiding relation is just the mathematical statement exactly as in Einstein's general theory of relativity. In Bohmian quantum gravity, particle trajectory is determined by the conformally flat space-time manifold $g_{\mu\nu}=\Omega^2\eta_{\mu\nu}$.
Klein-Gordon equation can be mapped to the conformally flat general relativity and the curved space-time structure is completely determined by the conformal factor $\Omega^2$. 

Still these arguments are not enough to see the full 
gravitational modification of quantum theory. One needs to couple gravity with Klein-Gordon matter field to see the whole correction. In order to couple gravitational field with quantum mechanical matter field, one is compelled to use a Lagrangian multiplier $\lambda$, which is essential for a such a theory.

\section{Coupling gravitational field with the quantum mechanical scalar matter field}
Following the work of Shojai et al.~\cite{Shojai_Article,ShojaiPRD99} and others ~\cite{GabayJoseph1,GabayJoseph2,SKJoseph1,SKJoseph2}, it had been shown that taking the following 
action,
\begin{eqnarray}
A[g_{\mu\nu},{\Omega}, S, \rho, \lambda]&=
&\frac{1}{2k}\int{d^4x\sqrt{-g}\left(R\Omega^2-6\nabla_{\mu}\Omega\nabla^{\mu}\Omega\right)}\nonumber\\
& &+\int{d^4x\sqrt{-g} \left(\frac{\rho}{m}\Omega^2 \nabla_{\mu}{S}\nabla^{\mu}{S}-m\rho\Omega^4\right)} 
\nonumber\\
& &+\int{d^4x\sqrt{-g}\lambda 
	\left[\ln(\Omega^2)-\left(\frac{\hbar^2}{m^2}\frac{\nabla_{\mu}\nabla^{\mu}\sqrt{\rho}}{\sqrt{\rho}}
	\right)\right]} \label{action}
\end{eqnarray}
%+\frac{\beta}{2}\,\Omega^{4}\,\theta \, {F}_{\mu\nu} \widetilde{F}^{\mu\nu}
and using Euler-Lagrange equation, quantum mechanical equation of motion for $\rho$ and $S$ can be found.  
Minimizing the action with respect to $\rho$ and $S$ 
leads to real and imaginary parts of the Generalized Klein-Gordon Equation. 
Note that the real part of the Klein-Gordon equation obeys the following equation 
\begin{equation}
\begin{split}
\nabla_{\mu}S \nabla^{\mu}S-m^2\Omega^2+\frac{\hbar^2}{2m 
\Omega^2\sqrt{\rho}}\Bigl[\Box{\Bigl(\frac{\lambda}{\sqrt{\rho}}\Bigr)}-\lambda\frac{\Box\sqrt{\rho}}
{\rho}\Bigr]=0. \label{EqMotion} 
\end{split}
\end{equation}
while the imaginary part gives the generalized continuity equation
\begin{eqnarray}
\nabla_{\mu}(\rho\Omega^2\nabla^{\mu}S)=0 \label{ContiEqn}.
\end{eqnarray}
Note that  Eq.~\ref{EqMotion} and Eq.~\ref{ContiEqn} represents quantum-mechanical wave-equations in a 
pure geometrical manner.
Here probabilities doesn't arises at all. Interpretationally things are comparably simpler. Here the 
particle obeys a more general equation of motion and continuity equation (Eq.~\ref{EqMotion} and Eq.~
\ref{ContiEqn} respectively).

Combining these two equations (Eq.~\ref{EqMotion} and Eq.~\ref{ContiEqn}) into a single complex wave 
equation, the generalized Klein-Gordon equation can be recovered. Then one need to take the probability 
based interpretation of quantum mechanics. In the wavefunction picture it is nice to see that how a linear 
partial differential equation gets modified due to the gravitational corrections.

\section{Bohmian-Quantum-Gravity generalizes de Broglie-Bohm theory}
Apart from Eq.~\ref{EqMotion} and Eq.~\ref{ContiEqn}, there is an extra $\lambda$ field equation in the theory. From Refs.~\cite{GabayJoseph1,GabayJoseph2,SKJoseph1,SKJoseph2,SKJoseph1,SKJ_ScalarGWave2019}, it is known that Lagrange multiplier field $\lambda$ obeys the following interesting equation,
%\begin{eqnarray}
%& & \nabla_{\mu}S \nabla^{\mu}S-m^2\Omega^2+\frac{\hbar^2}{2m 
%	\Omega^2\sqrt{\rho}}\Bigl[\Box{\Bigl(\frac{\lambda}{\sqrt{\rho}}\Bigr)}-\lambda\frac{\Box\sqrt{\rho}}
%	{\rho}\Bigr]=0. \nonumber \\ \label{EqMotionGen} 
%\\& & \nabla_{\mu}(\rho\Omega^2\nabla^{\mu}S)=0 .\label{ContiEqGen}
%\end{eqnarray}

\begin{eqnarray}
\lambda=\frac{\hbar^2}{m^2(1-Q)}\nabla_{\mu}\Bigl(\lambda\frac{\nabla^{\mu}\sqrt{\rho}}{\sqrt{\rho}} 
\Bigr) \label{LambdaNiceEq1}
\end{eqnarray}
This seemingly simple extra equation makes things complicated. Before exploring the physical meaning of $\lambda$, one can look a much simpler situation, ie. Bohmian equation of motion and continuity equation (Eq.~\ref{EqMotion} and Eq.~\ref{ContiEqn}), without 
Lagrangian multiplier contributions (taking $\lambda=0$ or $\lambda=\rho$).
These  equations are  already explored in the previous works~
\cite{Shojai_Article,GabayJoseph1,GabayJoseph2},
\begin{eqnarray}
\nabla_{\mu}S\nabla^{\mu}S - m^2\Omega^2=0 \label{EoMnoVac}\\
\nabla_{\mu}(\rho\Omega^2\nabla^{\mu}S)=0 \label{EoMcnteqn}.
\end{eqnarray}
Here, $\Omega^2=\exp(\frac{\hbar^2}{m^2} \frac{\nabla_{\mu} \nabla^{\mu}\sqrt{\rho}}{\sqrt{\rho}})=\exp(Q)
$, where $Q=\frac{\hbar^2}{m^2} \frac{\nabla_{\mu} \nabla^{\mu}\sqrt{\rho}}{\sqrt{\rho}}$, is the Klein-
Gordon quantum potential. Just to see the modification of quantum mechanical wave-equation from these equations, one needs to do the reverse procedure, ie. the complex number formulation of the theory. 

\subsection{Gravitational correction as a dissipation in the extended quantum theory}
One can combine Eq.~\ref{EoMnoVac} and Eq.~\ref{EoMcnteqn} in order to obtain the modified Klein-Gordon 
Equation. 
Hence the generalized Klein-Gordon equation can be written as,
\begin{eqnarray}
\Box\phi+\frac{i}{\hbar}\Bigl(\frac{\nabla_{\mu}\Omega^2}{\Omega^2}{\nabla^{\mu}S}\Bigr)\phi+ \frac{m^2}
{\hbar^2}\phi=0. 
\label{WFeqn1} 
\end{eqnarray}
Equation~\ref{WFeqn1} is true for both cases, i.e. for the conformal factor $\Omega^2$ which is linear 
order in $Q$ ($\Omega^2=1+Q$) or to the exponential order case ($\Omega^2=\exp{(Q)}$). 
While the expression $\frac{\nabla_{\mu}\Omega^2}{\Omega^2}$ can be perceived as the quantum force for an 
exponential constraint 
($\Omega^2=e^{Q}$) of the conformal factor $\frac{\nabla_{\mu}\Omega^2}{\Omega^2}= \nabla_{\mu}
\ln{\Omega^2}=\nabla_{\mu}Q$. Hence,
\begin{eqnarray}
\Box\phi+\frac{i}{\hbar}\Bigl({\nabla_{\mu}Q\,\nabla^{\mu}S}\Bigr)\phi+ \frac{m^2}{\hbar^2}\phi=0. 
\label{WFeqn2} 
\end{eqnarray}
It is pretty straight forward to see that Eq.~\ref{EoMnoVac}, Eq.~\ref{EoMcnteqn} and $\Omega^2=\exp(Q)$ 
can give Eq.~\ref{WFeqn2}. In the usual Klein-Gordon equation, the continuity equation is given by $
\nabla_{\mu}(\rho\nabla^{\mu}S)=0$ while the conformal gravity action (see Eq.~\ref{action})  
generalizes the usual Klein-Gordon continuity equation $\nabla_{\mu}(\rho\nabla^{\mu}S)=0 \to \nabla_{\mu}
(\rho\Omega^2\nabla^{\mu}S)=0$ with an extra factor $\Omega^2$. Due to the presence of the conformal 
factor $\Omega^2$ in the continuity equation, an additional imaginary term appears as the conformal 
gravity correction to the Klein-Gordon equation. One can see that the  middle term $+\frac{i}{\hbar}
\Bigl({\nabla_{\mu}Q\,\nabla^{\mu}S}\Bigr)\phi$ arises as a conformal gravitational correction to the 
Klein-Gordon equation and it is interesting to note that such a correction appears as a dissipative 
contribution to the wave-equation ($\frac{i}{\hbar}({\nabla_{\mu}Q\,\nabla^{\mu}S})\phi$). But this 
equation is not complete, in order to have an equilibrium in the system one needs to think about a 
fluctuation contribution. One can interpret this flucutation as a quantum-vacuum contribution as well, 
see Ref.~\cite{GabayJoseph1,GabayJoseph2} for more details.

\subsection{Vacuum source and dissipation in extended quantum theory}
From the theoretical understanding of the fluctuation dissipation theorem ~\cite{Kubo_1966}, one can speculate that 
the dissipation should be followed by some fluctuation in the system. For example, according the 
fluctuation-dissipation theorem, if dissipation exists in a system, there is a reverse process 
related to fluctuations. 
%There are several examples which demonstrate fluctuation-dissipation 
%theorem, for example  drag of a particle inside a fluid has a corresponding fluctuation associated to it, 
%which is just the Brownian motion. Similarly, when an object absorbs light, it is followed by a thermal %radiation which is the fluctuation part associated to it. 
In the same manner, the gravitational conformal factor correction gives a dissipative contribution in the  wave equation while the fluctuation part is still absent, hence we need to take into account $\lambda$ contribution. Taking into account the full coupled equations, one needs to consider two real field equations (see Eq.~\ref{EqMotion}, Eq.~\ref {ContiEqn}) and also the $\lambda$ equation (Eq.~\ref{LambdaNiceEq1}).
Wave-function equation can be defined in the same manner as before. After substituting Eq.~\ref{ContiEqn} into Eq.~\ref{EqMotion}, more general wave-function equation coupled to the Lagrange multiplier field
$\lambda$ can be obtained, 
\begin{eqnarray}
\Box\phi+\frac{i}{\hbar}\Bigl(\frac{\nabla_{\mu}\Omega^2}{\Omega^2}{\nabla^{\mu}S}\Bigr)\phi+ \frac{m^2}{\hbar^2}\phi
= \Bigl[\frac{1}{2m\Omega^2\rho}\Bigl(\Box
- \frac{2m^2(1-Q)}{\hbar^2}\Bigr)\lambda\Bigr]\phi  \label{WFeqnVacuumF1}. 
\end{eqnarray}
The Lagrange multiplier field $\lambda$ obeys a first order differential equation coupled to  the density via $\sqrt{\rho}$ 
\begin{eqnarray}
\lambda=\frac{\hbar^2}{m^2(1-Q)}\nabla_{\mu}\Bigl(\lambda\frac{\nabla^{\mu}\sqrt{\rho}}{\sqrt{\rho}} \Bigr) \label{LambdaNiceEq1}
\end{eqnarray}
Eq.~\ref{WFeqnVacuumF1} and Eq.~\ref{LambdaNiceEq1} together yields, quantum gravity corrected Klein-Gordon equation with the Lagrange multiplier contribution. Since $\lambda$ contribution gives a fluctuation source, it can be interpreted as a vacuum field density. Quantum system is no longer closed one, it interacts with an extra vacuum field density $\lambda$.

Coupled Eq.~\ref{WFeqnVacuumF1} and Eq.~\ref{LambdaNiceEq1}  can be seen as a complete quantum mechanical extension of a linear partial differential equation into the nonlinear regime. In a dynamical system point of view, the equation is complete, it contains a dissipative contribution, in order to balance that a vacuum field source contribution $\lambda$ is also present. Seemingly simple but complicated $\lambda$ equation is coupled to the quantum mechanical wave equation. When $\lambda=\rho$ ie. the vacuum field density is equal to the quantum mechanical particle density, then the system achieves an equilibrium condition, the linear type of Klein-Gordon equation can be recovered. This is an interesting feature that the quantum-gravity consideration gives interesting correction terms to the usual quantum mechanical formulation. As speculated earlier, quantum theory can be recovered with an equilibrium condition $\lambda=\rho$. Hence the linear quantum theory is just an approximation of a much richer nonlinear partial differential equation.   

\section{Further Directions}
From the philosophical perspective of the dynamical system theory, one can argue that the usual quantum 
theory needs to be generalized into the nonlinear regime. These generalizations are not unique, there are 
several ways to extend quantum theory to the nonlinear regime. Here from the fundamental assumptions of 
general theory of relativity and the  quantum physics one needs to widen the mathematical possibilities of these theories and found a reasonable theoretical framework to merge it consistently. 

\begin{figure}[]
\center
\includegraphics[scale=0.4]{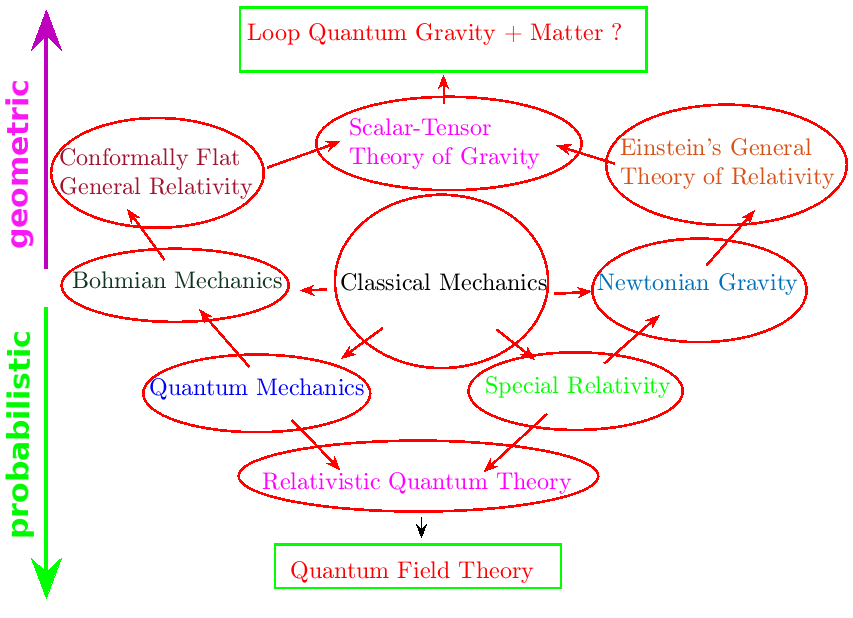}
\caption{Diagrammatic description of the proposed quantum-gravity approach using Bohmian version of 
quantum mechanics is shown. In Bohmian description, it is possible to geometrize quantum mechanics and 
incorporate it into the loop quantum gravity approach.}
\label{fig:1}
\end{figure}

From Fig.~\ref{fig:1}, one can obtain an overall idea of the theoretical scheme proposed here.
Figure~\ref{fig:1} shows two routes to unify quantum theory with general theory of relativity. 
Either we can adopt a probabilistic approach or a geometric approach.
If one thinks about probabilistic approach and its generalizations, then one needs 
to formulate general theory of relativity in a purely flat space-time background 
which we don't consider here. 
The other route is a purely geometric one (see Fig.~\ref{fig:1}), then we need to geometrize quantum mechanics and merge it general theory of relativity. Then both theories are geometric and dealing with them is rather simplified and all the interpretational issues of the orthodox quantum theory goes away.
Once quantum-mechanical matter field is incorporated into general relativity, one enters into the regime of the extended general theory of relativity  called the scalar-tensor theory~\cite{Shojai_ScalarTensor,Shojai2008,Fujii2003}.

Now one can think about quantizing this extended theory of gravity and it should eliminate different conceptual problems in the canonical quantum gravity. Gravitational field is still treated here as a classical background field and its full quantum properties are missing. In order to explore its quantum nature of gravity one needs to adopt the proper quantization schemes as used in canonical quantum gravity~\cite{DeWitt1967_I} or loop quantum gravity
~\cite{Ashtekar1986}. One of the biggest challenges in canonical quantum gravity is eliminating the problem of time. In authors perspective, such problems arise due to two reasons, (a) Einstein's general  theory of relativity is quantized via Wheeler-de-Witt equation, one needs to take into 
account the extended versions of general theory of relativity (b) Quantization procedure of Wheeler-de-
Witt equation is based on the simple functional Schr\"odinger equation. Rethinking and fixing (a) and (b), one should  be able to obtain the logically consistent theory of quantum-gravity.  In order to pursue such a scheme one needs to take a route of geometrizing quantum theory via conformally flat general relativity  presented here and then go to the scalar-tensor theory of gravity, then finally complete the quantization procedure using the extended quantum theory explored here.
 
\section{Conclusions}
It may take quite a long time to fully validate the theoretical aspects of a correct quantum-gravity 
theory. Yet the theory explored here, can be used as an intermediate testing ground. 
For example, one can look into the 
predictions coming from the nonlinear quantum theory. More importantly, this is a different route to quantum gravity, which is purely based on  a dynamical system point of view, which hints an entirely different way of extending both quantum theory and general relativity. Gravity corrects quantum theory to the nonlinear regime while quantum theory corrects general relativity to the scalar-tensor theory.
One needs to accept that, to get a fully consistent quantum-gravity theory both these well established
theories needs to be generalized.

%\section*{Acknowledgments}

%Inclusion of bibliography
%%%%%%%%%%%%%%%%%%%%%%%%%%%%%%%%
\halfnormalsize
\bibliographystyle{acm} 	%% The bibliography style file 'acm.bst' must be available in the folder.
\bibliography{MAT235_Joseph.bib} 	%% A file dsta-2019.bib must be available in the folder.
%							%% Use \cite{} to inclue a reference in the text.
%%%%%%%%%%%%%%%%%%%%%%%%%%%%%%%%%%%

% AUTHOR STAMPS AT THE END MUST BE INSERTED
\stamp{Sijo  K. Joseph}{Ph.D.}{WellGreen Technologies Pvt. Ltd.}{Thodupuzha, Kerala, India}{sijo@wellgreen.co.in}{}

%end
%
\end{document}